\documentstyle[12pt]{article}

\newcommand{\sgm}{\sigma}
\newcommand{\al}{\alpha}
\newcommand{\bt}{\beta}
\newcommand{\gm}{\gamma}
\newcommand{\Gm}{\Gamma}
\newcommand{\vp}{\varphi}
\newcommand{\ep}{\varepsilon}
\newcommand{\Om}{\Omega}
\newcommand{\om}{\omega}
\newcommand{\hH}{\hat H}
\newcommand{\be}{\begin{equation}}
\newcommand{\ee}{\end{equation}}
\newcommand{\Del}{\Delta}
\newcommand{\dlt}{\delta}
\newcommand{\lbd}{\lambda}
\newcommand{\ra}{\rightarrow}
\newcommand{\sr}{\stackrel{\rightarrow}{r}}
\newcommand{\sk}{\stackrel{\rightarrow}{k}}
\newcommand{\sE}{\stackrel{\rightarrow}{E}}
\newcommand{\sH}{\stackrel{\rightarrow}{H}}
\newcommand{\sD}{\stackrel{\rightarrow}{D}}
\newcommand{\sJ}{\stackrel{\rightarrow}{J}}
\newcommand{\sd}{\stackrel{\rightarrow}{d}}
\newcommand{\sA}{\stackrel{\rightarrow}{A}}
\newcommand{\ssp}{\stackrel{\rightarrow}{p}}
\newcommand{\prt}{\partial}
\newcommand{\su}{\stackrel{\rightarrow}{u}}
\newcommand{\snab}{\stackrel{\rightarrow}{\nabla}}

\begin{document}

\begin{center}
{\Large{\bf Coherent Polariton Radiation and Light Localization} \\ [5mm]
V.I. Yukalov} \\ [3mm]

{\it Centre for Interdisciplinary Studies in Chemical Physics \\
University of Western Ontario, London, Ontario N6A 3K7, Canada \\
and \\
Bogolubov Laboratory of Theoretical Physics \\
Joint Institute for Nuclear Research, Dubna 141980, Russia}

\end{center}

\vspace{2cm}

\begin{abstract}

A system of resonance atoms is placed in a medium with developed polariton
effect. In the spectrum of polariton states there can exist a band gap. If
an atom with a resonance frequency inside the polariton gap is incorporated 
into the medium, the atomic spontaneous emission can be strongly suppressed.
This is what is called the localization of light. Nevertheless, an ensemble 
of resonance atoms inside the polariton gap can radiate if their coherent
interaction is sufficiently strong. Conditions when coherent polariton
radiation can appear and the properties of this coherent polariton emission 
are studied.

\end{abstract}

\newpage

\section{Light Localization}

The phenomenon of light localization appears in three--dimensional
periodic dielectric structures, in which, due to periodicity, an
electromagnetic band gap develops. Then spontaneous emission with a
frequency inside the band gap can be rigorously forbidden [1--3], because
of a severe depression in the photon density of states for those
frequencies which remain in the spectral gap between the upper and lower
branches. This kind of samples, in which photon band gap develops because
of the structure periodicity in real space, has been called photonic
band--gap materials. The appearance of gaps in the spectrum of photon
states, due to real--space periodicity, is similar to the formation of
gaps in the spectrum of electrons in a periodic lattice potential [4].

If resonance atoms are incorporated in a band--gap material, so that their
transition frequency is inside the gap, then the effect of light
localization [2,3] can arize. To formulate explicitly what this effect
means, let us consider the average
$$
s(t) \equiv < \sgm^z(t) >
$$
of the population--difference operator $\sgm^z$. The average here implies
the statistical or, at zero temperature, quantum--mechanical average.
Under the localization of light one understands [5,6] that
$$
\lim_{t\ra\infty} s(t) > - 1\; .
$$

The light localization becomes possible due to the formation of
photon--atom bound states [5--7]. If a collective of identical impurity
atoms is incorporated into a medium with a photon band gap, so that their
transition frequency is inside the gap, and their spacing is much less
than the transition wavelength, then a photonic impurity band is formed
within the photonic band gap [7]. Electromagnetic coupling of neighboring
atoms takes place by means of an effective resonance dipole--dipole
interaction. If this interaction is sufficiently strong, then
electromagnetic radiation can propagate inside the impurity band [7].

The formation of photon band gaps in photonic band--gap materials is
similar to the well known polariton effect of the appearance of photon
bands due to the interaction of light with collective excitations of dense
medium [8,9]. Physical processes are, actually, the same in both types of
materials. The difference is only in the nature of scatterers which light
interacts with. In artificial photonic band--gap materials, a suppression
of the photon density of states over a narrow frequency range results from
multiple photon scattering by spatially correlated scatterers. In natural
dense media, such as dielectrics or semiconductors, a frequency gap for
propagating electromagnetic modes develops as a result of the photon
interaction with optical collective excitations, such as optical phonons,
magnons, excitons, and so on. Photons in a medium, coupled with collective
excitations, are called polaritons.

When a single resonance atom is placed in a frequency dispersive medium
whose polariton spectrum has a gap, and the atomic transition frequency
lies inside this gap, then a polariton--atom bound state appears causing a
significant suppression of spontaneous emission [10,11]. The physical
picture explaining this suppression is as follows. Let us imagine an atom
in a medium, with the atomic transition frequency within the polariton
gap. If this atom is initially excited, then it tends to become deexcited
emitting a photon. However, since the propagation of photons inside the 
polariton gap is prohibited, the emitted photon is scattered, by
collective excitations, back and is again absorbed by the atom. Thus, the
atom cannot get rid of a photon and is doomed to stay excited. Conversely,
if the atom is initially in the ground state, it continues to be in that
state, since there are no photons around to excite it. The supression of
spontaneous emission of an atom is termed localization of light. As is
explained above, the effect of light localization can be expressed as the
inequality $s(t)>-1$ for the average population difference, valid for all
times.

The formation of polariton--atom bound states has been studies for a
stationary case [10,11]. In dynamical picture, the population difference
of an atom satisfies the equation
$$
\frac{ds}{dt} = -\gm_1( s - \zeta ) \; ,
$$
in which $\gm_1$ is a level width and $\zeta$, a stationary value of the
population difference defined by a solution to the stationary problem.
From the dynamical equation one has
$$
s(t) = ( s_0 - \zeta ) e^{-\gm_1t} + \zeta \; ,
$$
where $s_0\equiv s(0)$. If the stationary value $\zeta>-1$, then
$\lim_{t\ra\infty}s(t) =\zeta>-1$, which implies the localization of
light. The complete suppression of emission corresponds to $\zeta=s_0$;
then $s(t)=\zeta$. Note that the linewidth $\gm_1$ is caused by vacuum
quantum fluctuations and is always nonzero, irrespectively what medium the
atom is placed into.

If a collection of resonance impurity atoms is doped into a medium with a
polariton band gap, then, in the same way as for photonic band--gap
materials [7], an impurity band can be formed within the polariton gap
[12,13]. Then electromagnetic radiation can propagate in such an impurity
band. In order that such an impurity band be formed, the spacing of
resonance impurity atoms in the medium should be much smaller than the
radiation wavelength. If it is so, then for a group of atoms a
sufficiently strong effective interaction, caused by photon exchange, can
develop. This interaction collectivizes the atoms that start radiating
coherently [14]. In this way, the suppression of spontaneous emission for
a single atom can be overcome by a group of atoms radiating coherently.

The situation when a single atom cannot radiate inside the polariton gap
but a collective of strongly interacting atoms can radiate reminds the
following related case. If a sample with a polariton band gap is
irradiated by a monochromatic electromagnetic wave with a frequency within
the polariton gap, then the incident light cannot propagate through this
medium because of total reflection. However, if the incident intensity is
large enough, the light can penetrate into the dense media even when
propagating inside the polariton gap [15,16]. In such a case, analogously
to that of coherent radiation, the possibility of the radiation
propagation inside the band gap is due to nonlinear effects.

Here and in what follows we use the term "atom" in the general sense,
implying under a resonance atom any two--level object. Depending on
radiation frequencies, this could be atoms as such, molecules, nuclei, or
quantum dots and wells. The latter case is of special importance for
semiconductors. Really, the polariton effect is well developed in many
semiconductors, for instance, in $CuCl,\; CuBr,\; CdSe,\; ZnSe,\; GaAs,\;
GaSb,\; InAs,\; AlAs,\; SiC$. The characteristic frequencies, where the
polariton band gap arises, are as follows [4] (see also [12--14]). For
example, in $GaAs$ the polariton gap having the width
$\Del\equiv\Om_2-\Om_1=4\times 10^{12}s^{-1}$ lies between
$\Om_1=5.1\times 10^{13}s^{-1}$ and $\Om_2=5.5\times 10^{13}s^{-1}$; the
linewidth being $\gm_1/\Om_1=1.2\times 10^{-5}$. In $SiC$ the polariton
gap $\Del=3\times 10^{13}s^{-1}$ is between $\Om_1=1.5\times
10^{14}s^{-1}$ and $\Om_2=1.8\times 10^{14}s^{-1}$; with the linewidth
$\gm_1/\Om_1=3\times 10^{-6}$. In all cases, for the relaxation parameter
$\gm_2$ one has $\gm_2/\Om_1\sim 10^{-2}$. As is seen, the polariton band
gap in such semiconductors is located in the infrared region. Therefore,
resonance radiation for this region of frequencies could be presented by
quantum dots and wells. Keeping in mind the feasibility of different
radiating objects, we continue, for the sake of simplicity, to use the
term "resonance atoms".

\section{Basic Equations}

The total Hamiltonian is given by the sum 
\be
\hH = \hH_a + \hH_f + \hH_m + \hH_{af} + \hH_{mf} \; ,
\ee
consisting of atomic, $\hH_a$, field, $\hH_f$, matter, $\hH_m$,
atom--field, $\hH_{af}$, and matter--field, $\hH_{mf}$, Hamiltonians. In
the atomic Hamiltonian
\be
\hH_a =\frac{1}{2} \sum_{i=1}^n \om_0 ( 1 +\sgm_i^z ) 
\ee
the index $i$ enumerates the atoms, $\om_0$ is a transition frequency, and
$\sgm_i^z$ is a population difference operator. Here and in what follows
we set $\hbar\equiv 1$. The field Hamiltonian
\be
\hH_f = \frac{1}{8\pi} \int \left [ \sE^2(\sr) + \sH^2(\sr) \right ] d\sr
\ee
has the standard form in which $\sE$ is electric field and
$\sH=\snab\times\sA$ is magnetic field, with a vector potential $\sA$
satisfying the Coulomb gauge condition $\snab\cdot\sA=0$. The Hamiltonian
of matter represents optic collective excitations and can be modelled by
an ensemble of oscillators,
\be
\hH_m =\sum_{i=1}^{N'} \frac{\ssp_i^2}{2m} + \frac{1}{2}\sum_{ij}^{N'}
\sum_{\al\bt}^3 D^{\al\bt}_{ij} u_i^\al u_j^\bt \; ,
\ee
where the index $i=1,2,\ldots,N'$ enumerate lattice sites, $\ssp_i$ and
$\su_i$ are momentum and displacement operators, respectively, and
$D_{ij}^{\al\bt}$ is a dynamical matrix. The atom--field interaction is
described by the Hamiltonian
\be
\hH_{af} = -\frac{1}{c} \sum_{i=1}^N J_a(\sr_i)\sA(\sr_i) \; ,
\ee
which corresponds to the dipole approximation with the transition current
\be
\sJ_a(\sr_i) = i\om_0\left ( \sgm_i^+ \sd^* -
\sgm_i^- \sd \right ) \; ,
\ee
where $\sgm_i^+$ and $\sgm_i^-$ are the rising and lowering operators,
respectively, and $\sd$ is a transition dipole. The matter--field
interaction can be presented as
\be
\hH_{mf} = -\frac{1}{c}\sum_{j=1}^{N'}\sJ_m(\sr_j)\sA(\sr_j) \; ,
\ee
with the matter current
\be
\sJ_m(\sr_j) = \frac{e}{m}\ssp_j \; ,
\ee
in which $e$ and $m$ are charge and mass, respectively.

The commutation relations for the operators introduced above are
$$
[ \sgm_i^+,\sgm_j^- ] =\dlt_{ij}\sgm_i^z \; , \qquad
[ \sgm_i^z,\sgm_j^\pm ] =\pm 2\dlt_{ij}\sgm_i^\pm \; ,
$$
$$
[ E^\al(\sr),A^\bt(\sr') ] = i4\pi c\dlt_{\al\bt}\dlt(\sr-\sr') \; .
$$
Using these relations and the Heisenberg equations of motion, we get the
Maxwell operator equations
\be
\frac{1}{c}\frac{\prt \sA}{\prt t} = -\sE \; , \qquad
\frac{1}{c}\frac{\prt \sE}{\prt t} =\snab\times \sH - \frac{4\pi}{c}\sJ \; ,
\ee
with the total density of current
\be
\sJ(\sr) =\sum_{i=1}^N \sJ_a(\sr_i) \dlt(\sr-\sr_i) +
\sum_{j=1}^{N'} J_m(\sr_j)\dlt(\sr-\sr_j) \; .
\ee
For the atomic variables, we find
\be
\frac{d\sgm^-}{dt} = - i\om_0\sgm_i^- + k_0\sgm_i^z \sd^*\cdot\sA_i\; ,
\ee
and
\be
\frac{d\sgm_i^z}{dt} = - 2k_0 ( \sgm_i^+ \sd^* + \sgm_i^- \sd)
\cdot \sA_i \; ,
\ee
where the notation
$$
\sA_i \equiv \sA(\sr_i,t) \; , \qquad k_0\equiv\frac{\om_0}{c}
$$
is used. From (9), with the Coulomb gauge condition, we have the wave
equation
\be
\left ( \snab^2 -\frac{1}{c^2}\frac{\prt^2}{\prt t^2}\right )\sA =
-\frac{4\pi}{c}\sJ\; ,
\ee
whose solution reads
\be
\sA(\sr,t) =\sA_v(\sr,t) +\frac{1}{c}\int
\frac{\sJ(\sr',t-|\sr-\sr'|/c)}{|\sr-\sr'|} d\sr'\; ,
\ee
where $\sA_v$, being a solution of the related homogeneous equation,
corresponds to vacuum fluctuations. Substituting the density of current
(10) into (14) yields for the vector potential at the point $\sr_i$ the 
expression
\be
\sA_i(t) = \sA_v(\sr_i,t) + \sA_a(\sr_i,t) + \sA_m(\sr_i,t) \; ,
\ee
in which the first term is caused by vacuum fluctuations, the second term,
\be
\sA_a(\sr_i,t) = ik_0\sum_{j(\neq i)}^N \frac{1}{r_{ij}}\left [
\sgm_j^+\left ( t -\frac{r_{ij}}{c}\right )\sd^* -
\sgm_j^-\left ( t -\frac{r_{ij}}{c}\right )\sd \right ] \; ,
\ee
where
$$
r_{ij} \equiv |\sr_{ij}| \; , \qquad \sr_{ij}\equiv \sr_i - \sr_j \; ,
$$
is a vector potential generated by radiating atoms, and the last term,
\be
\sA_m(\sr,t) = \frac{1}{c} \sum_{j(\neq i)}^{N'} \frac{1}{r_{ij}} \sJ_m
\left ( \sr_j, t -\frac{r_{ij}}{c}\right ) \; ,
\ee
is due to local electric currents in the medium. In the vector potentials
(16) and (17) the self--action parts are excluded. Instead, we shall add
to Eqs. (11) and (12) the terms describing the level width and the line
width,
$$
\gm_1 =\frac{2}{3} k_0^3 d_0^2 = \frac{1}{T_1} \; , \qquad
\gm_2 =\frac{1}{T_2}\; ,
$$
where $d_0\equiv |\sd|$. In this way, introducing the effective
electric induction
\be
\sD_i(t) \equiv k_0\left [ \sA_v(\sr,t) + \sA_m(\sr,t)\right ] \; ,
\ee
we come to the equations
$$
\frac{d\sgm_i^-}{dt} = - ( i\om_0 +\gm_2)\sgm_i^- +\sgm_i^z\sd^*\cdot
\sD_i +
$$
\be
+ ik_0^2\sgm_i^z\sd\cdot \sum_{j(\neq i)}^N \frac{1}{r_{ij}}\left [
\sgm_i^+\left ( t -\frac{r_{ij}}{c}\right )\sd^* -
\sgm_j^-\left ( t -\frac{r_{ij}}{c}\right ) \sd\right ]
\ee
and
$$
\frac{d\sgm_i^z}{dt} = -\gm_1 ( \sgm_i^z -\zeta )  + 2 (\sgm_i^z\sd^*
+ \sgm_i^- \sd) \cdot \sD_i -
$$
\be
- i2k_0^2( \sgm_i^+ \sd^* +\sgm_i^- \sd) \sum_{j(\neq i)}^N
\frac{1}{r_{ij}}\left [
\sgm_i^+\left ( t -\frac{r_{ij}}{c}\right )\sd^* -
\sgm_j^-\left ( t -\frac{r_{ij}}{c}\right ) \sd\right ] \; .
\ee

The retardation effects in these equations can be treated in the
quasirelativistic approximation. This means the following. In the
nonrelativistic limit, when $c\ra\infty$ and $k_0\ra 0$, from (19) would
follow $\sgm_i^-\sim \exp(-i\om_0 t)$. In the quasirelativistic
approximation, we set
\be
\sgm_i^-\left ( t -\frac{r_{ij}}{c}\right ) \simeq 
\sgm_i^-(t)\exp(ik_0r_{ij}) \; .
\ee

Define the statistical averages
\be
u_i\equiv < \sgm_i^- > \; , \qquad s_i \equiv < \sgm_i^z >
\ee
over atomic degrees of freedom. Then from (19) and (20), in the
semiclassical approximation, we obtain
\be
\frac{du_i}{dt} = - (i\om_0 +\gm_2 ) u_i + s_i (\sd_i^*\cdot \sD_i)
+ ik_0^3s_i\sd\cdot\sum_{j(\neq i)}^N (\vp_{ij}^*u_j^*\sd^* -
\vp_{ij}u_j\sd )
\ee
and
$$
\frac{ds_i}{dt} = -\gm_1 (s_i - \zeta ) - 2 (u_i^*\sd^* +
u_i\sd )\cdot \sD_i -
$$
\be
- i2k_0^3 ( u_i^*\sd^* + u_i\sd) \cdot
\sum_{j(\neq i)}^N ( \vp_{ij}^* u_j^*\sd^* -\vp_{ij} u_j\sd ) \; ,
\ee
where
$$
\vp_{ij} \equiv\frac{\exp(ik_0r_{ij})}{k_0r_{ij}} \; .
$$

The semiclassical approximation is a kind of the mean--field approximation. 
In the spirit of these, we may make the following mean--field approximation
$$
\sum_{j(\neq i)}^N \vp_{ij} u_j \approx u_i \sum_{j(\neq i)}\vp_{ij}
\equiv u_i\vp_i \; ,
$$
where
$$
\vp_i \equiv \sum_{j(\neq i)}^N \vp_{ij} =\sum_{j(\neq i)}^N
\frac{\exp(ik_0r_{ij})}{k_0r_{ij}}\; .
$$
The factors $\vp_{ij}$ and $\vp_i$ describe local fields.

Introduce the local--field shift
\be
\Del_L \equiv \gm_2g's \; , \qquad g' \equiv \frac{k_0^3d_0^2}{\gm_2}
\sum_{j(\neq i)}^N \frac{\cos(k_0r_{ij})}{k_0r_{ij}} \; ,
\ee
also called the cooperative Lamb shift [17], and the effective atom--atom
coupling parameter
\be
g \equiv \frac{k_0^3d_0^2}{\gm_2}\sum_{j(\neq i)}^N 
\frac{\sin(k_0r_{ij})}{k_0r_{ij}}\; .
\ee
These quantities enter into the definitions of the effective radiation
frequency and radiation width,
\be
\Om \equiv \om_0 +\Del_L \; , \qquad \Gm \equiv\gm_2( 1 - gs ) \; ,
\ee
respectively.

Involving these notations and keeping in mind that
$$
u_i = u(\sr_i,t) \; , \qquad s_i = s(\sr_i,t) \; , \qquad \sD_i =
\sD(\sr_i,t) , \qquad \vp_i = \vp(\sr_i) \; ,
$$ 
we transform Eqs. (23) and (24) to the form
\be
\frac{du}{dt} = - (i\Om +\Gm) u + s\sd^*\cdot \sD +
ik_0^3 s\vp^*u^*(\sd^*)^2
\ee
and
$$
\frac{ds}{dt} = - 4\gm_2 g |u|^2 -\gm_1 (s -\zeta ) - 2 (u^*\sd^*
+ u\sd )\cdot \sD -
$$
\be
- i2k_0^3 \left [ \vp^*(u^*\sd^*)^2 -\vp(u\sd)^2 \right ] \; .
\ee
Since $u$ is a complex variable, we have to supplement Eqs. (28) and (29)
by an equation for either $u^*$ or $|u|^2$. For instance, for $|u|^2$ we
get
$$
\frac{d|u|^2}{dt} = - 2\Gm |u|^2 + s (u^*\sd^* + u\sd)\cdot \sD +
$$
\be
+ ik_0^3 s\left [ \vp^*(u^*\sd^*)^2 -\vp(u\sd)^2\right ] \; .
\ee
Equations (29) and (30) give for the Bloch vector the equation
$$
\frac{d}{dt} \left ( s^2 + 4|u|^2\right ) = - 8\gm_2|u|^2 -
2\gm_1 ( s -\zeta )s \; .
$$
The derived equations (28), (29), and (30) are the basic equations
describing nonequilibrium processes in the system of resonance atoms
interacting with polariton field.

\section{Scale Separation}

Equations (28), (29), and (30) can be solved using the scale separation
approach [18--20]. To start with, we need to define what small parameters
we have.

The standard small parameters are related to the relaxation parameters
$\gm_1$ and $\gm_2$, for which
\be
\frac{\gm_1}{\om_0} \ll 1 \; , \qquad \frac{\gm_2}{\om_0}\ll 1\; .
\ee
It is reasonable to suppose that
\be
\left | \frac{\Del_L}{\Om}\right | < 1 \; , \qquad
\left | \frac{\Gm}{\Om}\right | < 1 \; ,
\ee
although $\Gm$ can become much larger than $\gm_2$. Assume also that
\be
\left | \frac{\sd\cdot\sD}{\Om}\right | < 1 \; ,
\ee
which means that the interaction of atoms with matter does not change
drastically the properties of the atoms. Under the validity of small
parameters (31) to (33), the variable $u$ has to be considered as fast,
compared to $s$ and $|u|^2$ that are to be treated as slow. Accepting the
variables $s$ and $|u|^2$ as quasi--integrals of motion, we keep them
fixed when solving Eq. (28). Then the solution for the fast variable is
$$
u(t) = u_0 G_1(t) + u_0^* G_2(t) +
$$
\be
+ s\sd^* \int_0^t\left [ G_1(t-\tau) + G_2(t-\tau)\right ]
\sD(\tau)d\tau \; ,
\ee
where $u_0\equiv u(0)$ and the Green functions are
$$
G_1(t) =\left (\frac{\lbd_1 -a^*}{\lbd_1-\lbd_2}\right ) e^{\lbd_1t} -
\left ( \frac{\lbd_2 -a^*}{\lbd_1-\lbd_2}\right ) e^{\lbd_2t} \; ,
$$
$$
G_2(t) =\frac{b}{\lbd_1-\lbd_2}\left ( e^{\lbd_1t} - 
e^{\lbd_2t}\right )\; ,
$$
$$
a = - (i\Om +\Gm ) \; , \qquad b = -i k_0^3s\vp^*(\sd^*)^2 \; ,
$$
$$
\lbd_{1,2} =\frac{1}{2} \left [ a + a^*\pm
\sqrt{(a-a^*)^2 + 4|b|^2}\right ] \; .
$$

Taking into account the existence of small parameters, we may write
$$
\lbd_{1,2} = \pm i\Om -\Gm \; ,
$$
and (34) can be simplified to
\be
u(t) = e^{-(i\Om+\Gm)t}\left [ u_0 + s\sd^*\int_0^t
e^{(i\Om+\Gm)\tau}\sD(\tau)d\tau \right ] \; .
\ee
The found fast variable (35) is to be substituted into the equations (29)
and (30) for the slow variables and the right--hand sides of these
equations are to be averaged over time and over the degrees of freedom
corresponding to collective excitations of matter [18--20]. Recall that
the quantities in (22) were defined as the averages over atomic degrees of
freedom. The double averaging, over time and over the matter degrees of
freedom, for a function $F(t)$, depending on time and on the operators of
collective excitations, is defined as
\be
<< F >> \equiv \lim_{\tau\ra\infty} \frac{1}{\tau} \int_0^\tau
< F(t)> dt \; ,
\ee
where the angle brackets imply the statistical averaging over the matter
degrees of freedom. The usage of the same angle brackets for denoting the
statistical averaging over the atomic and over matter degrees of freedom
should not yield confusion, since at the present stage the atomic degrees
of freedom do not arise being averaged out earlier. Therefore, in the
definition (36) and in what follows the statistical averaging always
concerns only the matter degrees of freedom.

Let us introduce the parameter
\be
\al \equiv <<\left | e^{-\Gm t}\int_0^t e^{(i\Om +\Gm)\tau}\sd^*\cdot
\sD(\tau) d\tau\right |^2 >>
\ee
characterizing the strength of interaction between the atoms and matter.
Thus, the quantity (37) can be called the atom--medium coupling parameter.

When substituting the fast variable (35) into the equations (29) and (30)
for the slow variables and averaging, according to (36), the right--hand
sides of the latter equations, we may notice the following useful
property. The fast variable (35) can be written as the sum $u=u_1+u_2$ of
the term
$$
u_1 = u_0 \exp\left\{ - (i\Om + \Gm) t\right \} \; ,
$$
not depending on the field $\sD$ of matter, and of the term
$$
u_2 = e^{-(i\Om+\Gm)t} s\sd^*\int_0^t e^{(i\Om+\Gm)\tau}\sD(\tau)d\tau\; ,
$$
depending on the matter field. Define the function
$$
w\equiv |u_1|^2 = |u_0|^2 e^{-2\Gm t} \; .
$$ By this definition, the function $w$ must satisfy an equation that
follows from the equation for $|u|^2$ where the matter field $\sD$ is set
zero. For $|u|^2$ we have
$$
|u|^2 = w + u_1^*u_2 + u_2^*u_1 + |u_2|^2 \; .
$$
When averaging, according to (36), we take into account that
$$ 
<< u_1^*u_2 + u_2^*u_1 >> = 0 \; .
$$
This is because of two reasons. First, the terms $u_1$ and $u_2$
oscillate, in general, with different frequencies. Second, the term $u_2$
is a linear combination of operators of collective excitations in matter.
In this way, averaging $|u|^2$ over fast variables, we get
$$
|u|^2 = w + << |u_2|^2 >> \; , \qquad
<< |u_2|^2 >> =\al s^2 \; .
$$
This consideration suggests that it is convenient to introduce the slow
variable
\be
w \equiv |u|^2 - \al s^2 \; ,
\ee
for which the evolution equation should have a form simpler than for
$|u|^2$. Really, averaging the equations (29) and (30) for the slow
variables, we obtain
\be
\frac{ds}{dt} = - 4g\gm_2 w - \gm_1 (s-\zeta) \; ,
\ee
where the transformation (38) is used, and
\be
\frac{dw}{dt} = - 2\gm_2 ( 1 -gs ) w\; .
\ee
From these two equations one can derive one equation
$$
\frac{d^2s}{dt^2} + ( 2 + \gm - 2gs )\frac{ds}{dt} - 2\gm g s^2 +
2\gm( 1 + g\zeta ) s - 2\gm\zeta = 0 \; ,
$$
in which $\gm=\gm_1/\gm_2$ and time is measured in units of $\gm_2^{-1}$.

\section{Models of Matter}

Before analysing equations (39) and (40), let us consider some examples
defining concretely the matter field $\sD$. Suppose, first, that the
matter consists of a set of random scatterers, such that
\be
\sd\cdot\sD = \xi \; ,
\ee
where $\xi$ is a stochastic field defined by the averages
\be
< \xi > = 0 \; , \qquad < |\xi|^2 > =\gm^2 \; .
\ee
Then the coupling parameter (37) is
\be
\al =\frac{\gm^2}{\Om^2} \; .
\ee

As another example, consider the matter modelled by the white noise, when
\be
\sd\cdot\sD =\xi(t) \; ,
\ee
where the white--noise stochastic variable $\xi(t)$ is defined by the
averages
\be
< \xi(t) > = 0\; , \qquad < \xi^*(t)\xi(t') > = 2\gm\dlt(t-t') \; ,
\ee
where the angle brackets mean a stochastic averaging. Then for the
coupling parameter, we get
\be
\al =\frac{\gm}{\Gm} \; .
\ee

In the third example, we model the matter by an oscillator, so that
\be
\sd\cdot\sD = \gm\left ( b_\om e^{-i\om t} + b_\om^\dagger e^{i\om t}
\right ) \; ,
\ee
where $b_\om$ and $b_\om^\dagger$ are the annihilation and creation
operators satisfying the Bose statistics, and for which the statictical
averaging gives
\be
< b_\om^\dagger b_\om > = n_\om \; , \qquad
< b_\om b_\om^\dagger > = 1 + n_\om \; ,
\ee
with $n_\om$ being an effective weight of excitations of a frequency
$\om$. Then the coupling parameter (37) is
\be
\al =|\gm|^2\left [ \frac{n_\om}{(\Om-\om)^2 +\Gm^2} +
\frac{1+n_\om}{(\Om+\om)^2 +\Gm^2}\right ] \; .
\ee
The strongest coupling between the impurity atoms and matter happens at
the resonance, when 
$\om = \Om$, and  $\al\cong n_\om|\gm/\Gm|^2$.

Finally, we consider a more realistic situation when the effective
electric induction of matter is defined by the relations (18), (17), and
(8), so that 
\be
\sD_i = \frac{ek_0}{mc} \sum_{j(\neq i)}^{N'} \frac{1}{r_{ij}}
\ssp_i\left ( t -\frac{r_{ij}}{c}\right ) \; ,
\ee
with the momentum operator
$$
\ssp_j(t) = - i\sum_{ks} \left ( \frac{m\om_{ks}}{2N'}\right )^{1/2}
\stackrel{\ra}{e}_{ks}\times
$$
\be
\times \left [ b_{ks}\exp\{ i(\sk\cdot\sr_j - \om_{ks}t)\} -
b_{ks}^\dagger\exp\{ - i(\sk\cdot\sr_j -\om_{ks}t)\} \right ] \; ,
\ee
in which $\om_{ks}=\om_{-ks}$ is a spectrum of collective excitations; 
$\sk$ being a wave vector; $s=1,2,3$, a polarization index; 
$\stackrel{\ra}{e}_{ks}$ is a polarization vector; $N'$ is the number of
lattice sites. The annihilation and creation operators of collective
excitations satisfy the Bose statistics and have the following statistical
averages
\be
< b_{ks}^\dagger b_{k's'} > = n_{ks}\dlt_{kk'}\dlt_{ss'} \; , \qquad
< b_{ks}b_{k's'} > = 0 \; .
\ee
In this case, for the coupling parameter (37), we obtain
\be
\al =\frac{k_0r_e}{2N'} \sum_{ks} f_{ks}\gm_{ks}\om_{ks} \left [
\frac{n_{ks}}{(\Om-\om_{ks})^2+\Gm^2} +
\frac{1+n_{ks}}{(\Om+\om_{ks})^2+\Gm^2}\right ] \; ,
\ee
where
\be
\gm_{ks}\equiv k_0^3|\sd\cdot\stackrel{\ra}{e}_{ks}|^2 \; , \qquad
r_e\equiv \frac{e^2}{mc^2} \; ,
\ee
and
\be
f_{ks} \equiv\left |\sum_{j(\neq i)}^{N'} 
\frac{\exp\left\{ i \left (\sk\cdot\sr_j +\frac{\om_{ks}}{c}r_{ij}\right )
\right\}} {k_0r_{ij}}\right |^2\; .
\ee
Again, it is clear that the coupling parameter (53) is the most strongly
influenced by resonance collective excitations with $\om_{ks}\approx\Om$.

\section{Transient Regime}

Consider the times shorter than $\gm^{-1}_1$. Then the term containing
$\gm_1$ in (39) can be omitted. In this case, using the second relation
from (27), we have
\be
\frac{d\Gm}{dt} = 4g^2\gm_2^2 w\; , \qquad
\frac{dw}{dt} = - 2\Gm w \; .
\ee
These two equations can be reduced to one,
$$
\frac{d^2\Gm}{dt^2} + 2\Gm\frac{d\Gm}{dt} = 0 \; , 
$$
integrating which we get
$$
\frac{d\Gm}{dt} + \Gm^2 = \gm_0^2 \; ,
$$
$\gm_0$ being an integration constant. The last equation is a Riccati
equation whose solution is
\be
\Gm = \gm_0 {\rm tanh}\left (\frac{t-t_0}{\tau_0}\right ) \; , \qquad
\gm_0 \equiv\frac{1}{\tau_0} \; ,
\ee
where $t_0$ is another integration constant. Using relation (27) gives
\be
s = -\frac{\gm_0}{g\gm_2}{\rm tanh}\left (\frac{t-t_0}{\tau_0}\right ) +
\frac{1}{g} \; .
\ee
The first equation in (56), together with (57), yields
\be
w=\frac{\gm_0^2}{4g^2\gm_2^2}{\rm sech}^2\left (
\frac{t-t_0}{\tau_0}\right ) \; .
\ee
And from (38) we find
\be
|u|^2 =\frac{\gm_0^2}{4g^2\gm_2^2}{\rm sech}^2\left (\frac{t-t_0}{\tau_0}
\right ) + \al s^2 \; .
\ee
The integration constants $\gm_0$ and $t_0$ are to be defined from the
initial conditions
\be
u(0) = u_0 \; , \qquad s(0) = s_0 \; .
\ee
From the latter we obtain for the effective radiation width $\gm_0$ the
expression
\be
\gm_0^2 = \Gm_0^2 + 4g^2\gm_2^2 \left ( |u_0|^2 -
\al_0 s_0^2\right ) \; ,
\ee
in which $\al_0$ is $\al$ at $t=0$, when $s=s_0$, 
$$
\Gm_0\equiv \gm_2 ( 1 - gs_0 )\; ,
$$
and the delay time
\be
t_0 =\frac{\tau_0}{2}\ln \left | \frac{\gm_0 - \Gm_0}{\gm_0 +\Gm_0}
\right | \; .
\ee
The radiation width can be written in the form
\be
\gm_0 = 2|gs_0|\gm_2 ( \al_c - \al_0 )^{1/2} \; ,
\ee
where the critical atom--matter coupling parameter
\be
\al_c \equiv \frac{(1-gs_0)^2 + 4g^2|u_0|^2}{4g^2s_0^2} 
\ee
is introduced. Expression (64) is a direct consequence of (62) for all $g$
and $\al_0$.

It is necessary to stress that the atom--matter coupling parameter (37)
cannot surpass the critical value (65). If this would happen, then the
radiation width (64) would become imaginary and, instead of (57), we would
have
$$
\Gm = - |\gm_0|{\rm tan}\left (\frac{t-t_0}{|\tau_0|}\right ) \; , \qquad
|\tau_0| \equiv\frac{1}{|\gm_0|} \; , 
$$
$$ 
|\gm_0| = 2\gm_2 |gs_0|\sqrt{\al_0 -\al_c} \qquad (\al_0 > \al_c)\; .
$$
The delay time (63) would be
$$
t_0 = |\tau_0|{\rm arctan}\left (
\frac{1-gs_0}{2|gs_0|\sqrt{\al_0-\al_c}}\right ) \; .
$$
And for the solutions (58) and (59) we would get
$$
s = s_0{\rm sgn}(gs_0)\sqrt{\al_0 -\al_c}{\rm tan}\left (
\frac{t-t_0}{\tau_0}\right ) +\frac{1}{g} \; ,
$$
$$
w = - s_0^2 (\al_0 - \al_c){\rm sec}^2\left ( 
\frac{t-t_0}{\tau_0}\right ) \; .
$$
The effective width $\Gm$, as well as the solutions $s$ and $w$, become
divergent at $t=t_n$,
$$
t_n = t_0 +\frac{\pi}{2} ( 1 + 2n) |\tau_0| \qquad (n=0,1,2\ldots) \; .
$$
Certainly, this behaviour is unphysical and it means that some conditions,
under which the method of scale separation has been used, are, probably,
not valid any more. This is really the case since, when $\Gm$ and $s$
diverge, conditions (32) do not hold true. Consequently, when
$\al_0>\al_c$, we cannot separate solutions onto fast and slow, all of
them oscillating equally fast. At the same time, the existence of slow
solutions is a characteristic feature of developed coherence. Thus, the
absence of slow solutions suggests that coherence cannot emerge in the
system. From the physical point of view, all this sounds quite
understandable. There should be a threshold for the strength of
interactions of atoms with matter after which such strong interactions
destroy the correlation between atoms, thus, destroying their coherence.
In this way, the inequality
\be
\al_0 < \al_c
\ee
is a necessary condition for the applicability of the scale separation
approach and, at the same time, a condition for the possibility of
coherent radiation of doped atoms.

The maximal level of coherence develops at the time $t=t_0$ when
\be
s(t_0) =\frac{1}{g}\; , \qquad w(t_0) = s_0^2 (\al_c -\al_0 )\; , \qquad
|u(t_0)|^2 = (\al_c -\al_0) s_0^2 + \frac{\al}{g^2} \; .
\ee
For the times much longer than $t_0$, Eqs. (58) to (60) give
$$
s\simeq \frac{1}{g} \left ( 1 -\frac{\gm_0}{\gm_2}\right ) \qquad
( t \gg t_0) \; ,
$$
\be
w \simeq \frac{\gm_0^2}{g^2\gm_2^2}\exp\left ( -2\gm_0 t\right ) \; ,
\ee
$$
|u|^2 \simeq \frac{\gm_0^2}{g^2\gm_2^2}\exp\left ( -2\gm_0 t\right ) +
\frac{\al}{g^2}\left ( 1 -\frac{\gm_0}{\gm_2}\right )^2 \; .
$$
However, the asymptotic behaviour given by (68) is valid only for $t\ll
T_1$.

Consider the case when both the atom--matter and atom--atom coupling
parameters are small, i.e.
\be
\al_0 \ll \al_c \; , \qquad |g| \ll 1 \; .
\ee
Using the first of inequalities in(69), we have from (64)
$$
\gm_0 \simeq \gm_2\left [ ( 1 - gs_0 )^2 + 4g^2 |u_0|^2 \right ]^{1/2} 
\left ( 1 -\frac{\al_0}{2\al_c}\right ) \; .
$$
The latter expression, with the second inequality in (69), reduces to
$$
\gm_0 \simeq \gm_2 (1 - g s_0)\left ( 1 -\frac{\al_0}{2\al_c}\right )\; .
$$
The critical parameter (65) becomes
\be
\al_c \simeq ( 4g^2 s_0^2)^{-1} \qquad ( g\ll 1 ) \; .
\ee
Employing this, we find
\be
\gm_0 \simeq \gm_2 ( 1 -gs_0 - 2 \al_0 g^2 s_0^2 ) \; ,
\ee
valid for small coupling parameters as in (69). For the delay time (63),
we get
\be
t_0 \simeq \frac{1+gs_0}{2\gm_2}\ln\left | \al_0 g^2 s_0^2\right |\; ,
\ee
which tends to $-\infty$ if either $\al_0$ or $g$ tends to zero. This
implies that, under conditions (69), an essential level of coherence does
not evolve.

Let us analyse the case when
\be
\al_0 \ll \al_c \;, \qquad |g| \gg 1\; .
\ee
Then the critical parameter (65) is
\be
\al_c \simeq \frac{1}{4s_0^2} \left ( s_0^2 + 4|u_0|^2 - 2\frac{s_0}{g}
\right ) \; .
\ee
The radiation width (64) becomes
\be
\gm_0 \simeq \frac{\gm_2|g|}{\sqrt{s_0^2 + 4|u_0|^2}}\left ( s_0^2
+ 4|u_0|^2 - 2\al_0 s_0^2 - \frac{s_0}{g} \right ) \; ,
\ee
with the corresponding radiation time
\be
\tau_0 \simeq \frac{T_2}{|g|\sqrt{s_0^2 + 4|u_0|^2}}\; .
\ee
For the delay time (63), we find
\be
t_0 \simeq \frac{\tau_0}{2}\ln\left |
\frac{|g|(s_0^2 + 4|u_0|^2-2\al_0s_0^2) + gs_0\sqrt{s_0^2+4|u_0|^2}}
{|g|(s_0^2 + 4|u_0|^2-2\al_0s_0^2) - gs_0\sqrt{s_0^2+4|u_0|^2}}\right |\; .
\ee
If the process develops from an initially incoherent state, when $u_0=0$,
then the radiation width (75) is
\be
\gm_0 \simeq \gm_2 |gs_0| \left ( 1 - 2\al_0 -\frac{1}{gs_0}\right )
\qquad (u_0 = 0 )\; .
\ee
Thence, the delay time (77) becomes
\be
t_0 \simeq\frac{T_2}{2|gs_0|}\ln\left | 
\frac{1-2\al_0 +\ep}{1-2\al_0-\ep}\right | \; ,
\ee
where
$$
\ep\equiv{\rm sgn}(gs_0) =\pm 1 \; .
$$
As far as for $u_0=0$ and $|g|\gg 1$, the critical parameter (74) is
$$
\al_c \simeq \frac{1}{4} \qquad (|g|\gg 1, \; u_0=0) \; ,
$$ then the inequality $\al_0\ll\al_c$ implies $\al_0\ll 1$. Hence, we may
simplify (79) as
\be
t_0 \simeq \frac{T_2}{2gs_0}\left |\ln\al_0\right | \; .
\ee
After the time (80), the population difference tends to
\be
s\simeq -\ep s_0( 1 -2\al_0) +\frac{1+\ep}{g} \qquad (t\gg t_0) \; .
\ee
If $g>0$, then $\ep s_0=|s_0|$, while for $g<0$, one has $\ep s_0=-|s_0|$.
Combining both these cases, we get $\ep s_0 ={\rm sgn}(g)|s_0|$.

Assume now that the atom--matter coupling parameter is close to its
critical value (65), but the atom--atom coupling is arbitrary,
\be
\frac{|\al_0-\al_c|}{\al_c} \ll 1 \qquad (\forall g) \; .
\ee
Then the radiation width (64) tends to zero, as $\al_0\ra\al_c$, and
respectively, the radiation time $\tau_0\equiv\gm_0^{-1}$ tends to
infinity. For the delay time (63), we find
\be
t_0\simeq \frac{2|gs_0|T_2}{(1-gs_0)^2} (\al_c -\al_0)^{1/2} \; ,
\ee
while the radiation time is
\be
\tau_0 =\frac{T_2}{2|gs_0|}(\al_c -\al_0)^{-1/2} \; .
\ee
When $\al_0\ra\al_c$, then $t_0\ra 0$, and for the functions (58) and (59)
we have
$$
s\simeq \frac{1}{g} - 2|s_0|{\rm sgn}(g)\sqrt{\al_c-\al_0}\left (
1 - 2e^{-2\gm_0t}\right ) \; ,
$$
\be
w\simeq 4s_0^2 (\al_c - \al_0 ) e^{-2\gm_0t} \qquad (t > t_0) \; .
\ee
There is a suppression of self--organized coherence in the system of
atoms, their radiation being almost completely due to the pumping by
matter excitations
$$
s \approx \frac{1}{g}\; , \qquad w\approx 0 \; , \qquad
|u|^2 \approx \frac{\al}{g^2} \; .
$$
Although the coherent relaxation may happen provided that $\tau_0\ll T_2$,
that is,
$$
|gs_0|(\al_c -\al_0)^{1/2} \gg 1 \; ,
$$
which corresponds to superradiant emission.

Note that analysing the properties of the solutions to equations (39) and
(40), we talk about radiation processes keeping in mind the following. The
total radiation intensity of atoms can be approximately defined in the
usual way as
\be
I(t) = - N\hbar \om_0\frac{ds}{dt} \; .
\ee
For a more accurate definition of radiation intensity see e.g.
Refs.[21,22]. From (86), using equation (39), we find
\be
I(t) = I_{coh}(t) + I_{inc}(t) \; ,
\ee
where the first term
\be
I_{coh}(t) = 4Ng\hbar\om_0\gm_2 w
\ee
had the meaning of the coherent radiation intensity, and the second,
\be
I_{inc}(t) = N\hbar\om_0\gm_1 (s -\zeta ) \; ,
\ee
corresponds to the intensity of incoherent radiation. The latter is always
proportional to the number of atoms $N$, while the radiation intensity
(88) is proportional to $Ng$. For a concentrated sample, whose linear size
is much smaller that the radiation wavelength, we have $g\approx N$, and
the radiation intensity (88) becomes proportional to $N^2$, which is
typical of superradiance. In this way, the solutions $s$ and $w$ define
the temporal behaviour of the incoherent radiation intensity (89) and of
the coherent radiation intensity (88), respectively. For instance, using
the solution $w$ given by (59), with the radiation width (64), we obtain
the intensity of coherent radiation
\be
I_{coh}(t) = 4Ng\hbar\om_0\gm_2 s_0^2 (\al_c -\al_0){\rm sech}^2\left (
\frac{t-t_0}{\tau_0}\right ) \; .
\ee
The latter shows that, if $\al_0\ra\al_c$, then $I_{coh}\ra 0$.

\section{Close--to--Stationary Regime}

In the previous section the transitient regime is considered corresponding
to times $t\ll T_1$. For the times comparable or larger that $T_1$, we
cannot neglect any more the term with $\gm_1$ in equation (39). In the
intermediate stage, when $t\sim T_1$, an exact solution of Eq.(39) and
(40) is not available. Here we have to resort to numerical calculations,
which will be the subject of a separate paper. But it is possible to give
an analysis for asymptotically large times, when $t\gg T_1$.

The following analysis assumes that $g\neq 0$. Since, if $g=0$, the
solutions to Eqs.(39) and (40) are
$$
s =\zeta + ( s_0 -\zeta ) e^{-\gm_1 t} \; ,
$$
\be
w = \left ( |u_0|^2 - \al s_0^2 \right ) e^{-2\gm_2 t} \qquad (g=0) \; ,
\ee
which describes the relaxation process of a single atom. In such a case,
if there is the localization of light, then $\zeta=s_0$, and (91) gives
$s=s_0$.

If $N$ impurity atoms are doped into the matter, then $g\neq 0$. The
resonance dipole--dipole interactions of a pair of atoms with a transition
frequency inside the photon gap have been studied in several works
[7,23,24]. The conclusion of these studies is that two closely spaced
atoms, with transition frequencies in the gap, interact with each other by
means of the virtual photon exchanges much in the same way as the atoms in
vacuum. That is, if the atoms are separated from each other by a spacing
much larger than the radiation wavelength, then each of them can be
considered as a single atom. If the transition frequency of such an atom
is inside the gap, then the phenomenon of light localization occurs.
However, if the atoms are close to each other, with a spacing mush smaller
than the radiation wavelength, then they practically do not experience the
existence of the gap [7,23,24].

Consider the close--to--stationary regime, when $t\gg T_1$. Equations (39)
and (40) can be written as 
\be
\frac{ds}{dt} = V_1 \; , \qquad \frac{dw}{dt} = V_2 \; ,
\ee
with the right--hand sides
$$
V_1 = - 4g\gm_2 w - \gm_1 ( s - \zeta ) \; ,
$$
\be
V_2 = -2\gm_2 ( 1 -gs ) w \; .
\ee
Stationary points, or fixed points for Eqs.(92), (93), are given by the
condition $V_1=V_2=0$. This yields two stationary points:
\be
s_1^* =\zeta\; , \qquad w_1^* = 0
\ee
and
\be
s_2^*=\frac{1}{g} \; , \qquad 
w_2^* = -\frac{\gm_1(1-g\zeta)}{4\gm_2g^2}\; .
\ee
The stability of these fixed points can be defined by the Lyapunov
analysis. To this end, we need to find the eigenvalues of the Jacobian 
matrix
\begin{eqnarray}
\hat J=\left [ \begin{array}{cc}
\frac{\prt V_1}{\prt s} & \frac{\prt V_1}{\prt w} \\
\\
\frac{\prt V_2}{\prt s} & \frac{\prt V_2}{\prt w}
\end{array} \right ] \; .
\end{eqnarray}
These eigenvalues are given by the expression
\be
\lbd^\pm = -\frac{1}{2}\left \{ \gm_1 +2\gm_2(1 - gs) \pm
\left [ \left ( \gm_1 - 2\gm_2(1-gs)\right )^2 - 32 \gm_2^2 g^2 w
\right ]^{1/2}\right \} \; .
\ee
Substituting here the values corresponding to the fixed points yields the
Lyapunov exponents. For the stationary point (94), we have
\be
\lbd_1^+ = -\gm_1 \; , \qquad \lbd_1^- = -2\gm_2(1 -g\zeta)\; ,
\ee
and for the stationary point (95), we find
\be
\lbd_2^\pm = -\frac{\gm_1}{2}\left\{ 1 \pm \left [ 1 +
8\frac{\gm_2}{\gm_1}(1 -g\zeta)\right ]^{1/2}\right \} \; .
\ee

The analysis of the Lyapunov exponents (98) and (99) shows that if
\be
g\zeta < 1 \; ,
\ee
then the fixed point (94) is a stable node, and the fixed point (95) is a
saddle point. When
\be
g\zeta = 1 \; ,
\ee
both fixed points merge together becoming neutral, since
$\lbd_1^-=\lbd_2^-=0$. In this case, the system of equations (39) and (40)
is structurally unstable. Equality (101) defines a bifurcation point. For
the interval
\be
1 < g\zeta \leq 1 +\frac{\gm_1}{8\gm_2}\; ,
\ee
the fixed point (94) is a saddle point, while that (95) is a stable node.
For all $g\zeta > 1$, the point (94) is a saddle point. If
\be
g\zeta > 1 + \frac{\gm_1}{8\gm_2} \; ,
\ee
the stationary point (95) becomes a stable focus, since the Lyapunov
exponents (99) take the form
\be
\lbd_2^\pm = -\frac{\gm_1}{2} \mp i\om_\infty \; ,
\ee
where
$$
\om_\infty \equiv \frac{\gm_1}{2}\left [ \frac{8\gm_2}{\gm_1}
(g\zeta -1 ) - 1 \right ]^{1/2} \; .
$$

Suppose that for a single atom there occurs the localization of light, so
that $\zeta=s_0$. If many resonant atoms are doped into matter, but their
interactions through the polariton exchange are not strong enough, so that
$gs_0<1$, then the light localization prevails. This means that the light
remains confined  in the vicinity of the atoms. The confinement of light
is demonstrated by the fact that the stationary point (94) is a stable
node with $s_1^*=s_0$. But if the resonant interaction between the atoms
is sufficiently strong, so that
\be
gs_0 > 1 \; ,
\ee
the deconfinement of light happens. Then the fixed point (95) becomes
stable, while the point (94) looses its stability. The deconfinement of
light is not complete, since $s_2^*=1/g < s_0$, but it is partial. The
portion of light that remains confined decreases with increasing $g$. The
qualitative change of the asymptotic behaviour of solutions to a system of
differential equations is called, in dynamical theory, the dynamical phase
transition. In our case, this happens if $g\zeta = 1$. The equality
$gs_0=1$ separates the regions where light is localized ($gs_0<1$) and
where it is deconfined ($gs_0>1$). Therefore, the dynamical phase
transition occurring at $gs_0=1$ corresponds to a transition that may be
called the {\it deconfinement of light} or {\it photon deconfinement}.

When the resonant interaction between atoms is so strong that inequality
(103) holds true, then the stable stationary point (95) is a focus. This
means that the solutions to the equations (39) and (40) display an
oscillatory regime of motion when approaching the stationary point (95).
Such an oscillatory motion is similar to that found for a concentrated
system with the resonant frequency near the edge of a photonic band gap
[5] and to that for two atoms with transition frequencies inside or
slightly outside a photonic band gap [24].

\section{Coupling Parameters}

There are several characteristic quantities defining the behaviour of the
system. These are the initial conditions $s_0\equiv s(0)$ and
$u_0\equiv u(0)$ and the coupling parameters $g,\; g'$, and $\al$. Below
we study the typical values of the latter.

Recall that the coupling parameters $g'$, defined in (25), and $g$, given
in (26), have appeared in the evolution equations when treating the
retardation effects in the quasirelativistic approximation (21). Without
the latter approximation, we should deal with the integral--type equations
[25]. Thus, the atom--atom coupling parameters $g'$ and $g$ describe the
retardation or local--field effects. The values of these parameters
essentially depend on the shape of the sample and on the spacing between
atoms [17].

Accepting the equality $\gm_1=2k_0^3d_0^2/3$, we have from (25) and (26)
$$
g=\frac{3\gm_1}{2\gm_2}\sum_{j(\neq i)}^N
\frac{\sin(k_0r_{ij})}{k_0r_{ij}}\; , \qquad g' =\frac{3\gm_1}{2\gm_2}
\sum_{j(\neq i)}^N\frac{\cos(k_0r_{ij})}{k_0r_{ij}}\; .
$$
If the radiation wavelength $\lbd=2\pi/k_0$ is much smaller than the mean
spacing, $a$, between the atoms, then the sums in $g$ and $g'$ can be of
any sign but with absolute values less than unity. As far as usually
$\gm_1\ll\gm_2$, the absolute values $|g|$ and $|g'|$ are small. If
$|g|\ll 1$ and $|g'|\ll 1$, the impurity atoms almost do not interact with
each other and their behaviour is practically the same as that of a
collection of single atoms.

In the opposite case, when $\lbd\gg a$, the sums in $g$ and $g'$ can be
estimated with a good approximation [17] as
$$
\sum_{j(\neq i)}^N \frac{\sin(k_0r_{ij})}{k_0r_{ij}}\approx
\sum_{j(\neq i)}^N \frac{\cos(k_0r_{ij})}{k_0r_{ij}}\approx \rho\lbd^3\; ,
$$
where $\rho\equiv N/V$ is the density of the doped atoms and it is assumed
that $\lbd$ is less than the linear sizes of the sample in all directions.
Then we have
$$
g\approx g' \approx \frac{3\gm_1}{2\gm_2}\rho\lbd^3 \; .
$$

The value of the atom--matter coupling parameter (37) essentially depends
on the peculiarity of the atom--matter interaction. As the models of
Section 4 show, one should expect that $\al\ll 1$. If the transition
frequency of the doped atoms lies outside the polariton band gap, then the
atom--matter resonance is possible, when $\om_{ks}\sim\om_0$. Moving the
atomic frequency into the gap makes such a resonance more and more
difficult. Far inside the gap, where there are no elementary excitations
of matter, this resonance becomes impossible. It follows from (53) that
the relation between the atom--matter coupling parameter, $\al_{out}$,
corresponding to the case when the atomic frequency is outside the gap,
and the parameter $\al_{ins}$, when the frequency is far inside the gap,
is roughly speaking, as
$$
\frac{\al_{out}}{\al_{ins}}\sim \frac{\Om^2}{\Gm^2}\; ,
$$
for a sufficiently large polariton band gap. The decrease of $\al$ leads,
as is clear from (80), to the increase of the delay time $t_0$.

One can also notice that the coupling parameters $g$ and $\al$ are not
independent, but $\al$ depends on $g$. This dependence, for $g\gg 1$, is
approximately as $\al\sim g^{-2}$. Therefore, strong atomic interactions
diminish $\al$, thus, increasing the delay time (80). During the interval
$0\leq t < t_0$, there is a temporal localization of light even for
rather large parameters $g$, such that $g\zeta > 1$, but then the process
of photon deconfinement starts. If $t_0$ becomes comparable with $T_1$,
one cannot omit in Eqs. (39), (40) the term containing $\gm_1$. Then one
should resort to numerical solution of these equations, which will be 
considered in a separate paper.

\vspace{5mm}

{\bf Acknowledgement}

\vspace{2mm}

I am grateful to M.R. Singh for useful discussions. Financial support from
the University of Western Ontario, Canada, is appreciated.

\end{document}